\documentclass{aa520}
\usepackage{graphicx,txfonts}

\def\kms{$\mathrm{km\;s}^{-1}$}

\def\ha{H$\alpha$}
\def\hb{H$\beta$}
\def\h2{H$_{2}$}
\def\hi{H~{\small I}}
\def\hii{H~{\small II}}
\def\nii{[N~{\small II}]}

\def\niig{[N~{\small II}]$\,\lambda6583$}
\def\niipg{[N~{\small II}]$\,\lambda\lambda6548,6583$}
\def\oiii{[O~{\small III}]$\,\lambda5007$}

\def\siipg{[S~{\small II}]$\,\lambda\lambda6716,6731$}

\begin{document}

\title{NGC~7679: an anomalous, composite Seyfert~1 galaxy whose,
X-ray luminous AGN vanishes at optical wavelengths.\thanks{Based
on observations obtained with the New Technology
Telescope at the European Southern Observatory (ESO 63.N-0024B)
in La Silla (Chile), with the Multiple Mirror Telescope---a joint
facility of the Smithsonian Institution and the University of
Arizona (USA)---and with the 1.82-m telescope at the Mount Ekar
Observatory in Asiago (Italy).}}

\author{L.M.~Buson\inst{1}, M.~Cappellari\inst{2},
E.~M.~Corsini\inst{3}, E.~V.~Held\inst{1},
J.~Lim\inst{4}, and A.~Pizzella\inst{3}}

\offprints{Lucio M. Buson,\\ email:{\tt buson@pd.astro.it}}

\institute{
INAF Osservatorio Astronomico di Padova, vicolo
dell'Osservatorio~5, I-35122 Padova, Italy\and
Leiden Observatory, Postbus 9513, 2300 RA Leiden, The Netherlands\and
Dipartimento di Astronomia, Universit\`a di Padova, vicolo
dell'Osservatorio~2, I-35122 Padova, Italy\and
Institute of Astronomy and Astrophysics, Academia Sinica,
P.O. Box 23-141, Taipei 106, Taiwan}

\date{Received ... / Accepted 24 September 2005 }

\titlerunning{NGC 7679}
\authorrunning{Buson et al.}

\abstract{Morphological disturbances and gas kinematics of the SB0
galaxy NGC~7679=Arp~216 are investigated to get clues to the history
of this highly composite object, where AGN and starburst signatures
dominate each other in the X-ray and optical/IR regime,
respectively. Perturbations of the ionized gas velocity field appear
quite mild within $15''$ ($\sim$5~kpc) from the center, so as it can
be straightforwardly modeled as a circularly rotating disk. On the
contrary, outside that radius, significant disturbances show up. In
particular, the eastern distorted arm as well as the huge neutral
hydrogen bridge connecting NGC~7679 with the nearby Seyfert spiral
NGC~7682 unambiguously represent the vestige of a close encounter of
the two objects dating back $\sim$500~Myr ago. The relationship of
such past event with the much more recent, centrally located starburst
(not older than 20~Myr) cannot be easily established. Altogether,
the classification of NGC~7679, turns out to be less extreme than that
proposed in the past, being simply a (disturbed) galaxy where
starburst and AGN activity cohexist with a starburst dominating the
bolometric luminosity.
}

\maketitle

\section{Introduction}

The barred galaxy we discuss in detail here, namely NGC~7679, has all
the necessary ``qualifications'' to be regarded as a rare member of a
class of composite (AGN/starburst) systems whose X-ray bright Seyfert
nucleus becomes unexpectedly weak at optical wavelengths. Though a few
objects of this kind are known since observations with the {\em
Einstein} spacecraft ({\em e.g.} Elvis et al. 1981) and are currently
identified in {\em Chandra} deep-fields ({\em e.g.} Fiore et
al. 2000), NGC~7679 could represent a {\em low-redshift} example of
such peculiar objects showing anomalous absorption processes, {\em
i.e.} hosting dusty ionized absorbers capable of {\em selectively}
obscure the AGN optical emission, while leaving its X-ray emission
almost unabsorbed ({\em cf.} Della Ceca et al. 2001).

Other well-known obscured Type-2 AGNs, namely NGC~4945 and
NGC~6240 (see Matt 2001 and references therein) are, at some extent,
reminiscent of the phenomenology showed by NGC~7679 ({\em e.g.} the
full emergence of the powerful AGN emission only in the X-ray
waveband). NGC~7679, however, is so peculiar as to challenge the
standard AGN Unification Scheme since -- presumably -- its absorption
processes can work quite differently, so as to leave its X-ray
emission {\em totally} unabsorbed, while producing a limited amount of
absorption of the AGN in the optical (or, alternatively, hosting an
optically underluminous Seyfert~1 nucleus; {\em cf. } Della Ceca et
al. 2001).

Simply classified SB0p in the Third Reference Catalogue of Bright
Galaxies (RC3; de Vaucouleurs et al. 1991), NGC~7679 appears
unmistakably disturbed, so as to have been included in the Arp's
catalog of peculiar galaxies (Arp 1966). In particular, its asymmetric
morphology points toward a barred Seyfert companion (NGC~7682), lying
only $\sim$ 4$'$.5 North-East of it. The recession velocity of
NGC~7682 ($V_{3K} = 4762\pm19$ km s$^{-1}$) is virtually identical to
that of NGC~7679 ($V_{3K} = 4778\pm13$ km s$^{-1}$), once both are
referred to the 3K background on the basis of their RC3 redshift and
cosmological parameters H$_0$=75 km s$^{-1}$ Mpc$^{-1}$ and
q$_0$=0. We adopt here the distance of NGC~7679 of 63.9~Mpc; at such a
distance 1$'$ is equivalent to $\sim$19~kpc and the projected distance
of the two galaxies becomes 83.7~kpc.

Strange enough, this object received only sporadic attention in the
literature in the past two decades. The optical spectroscopic survey
of Dahari (1985) includes NGC~7679 in the transition class of
\hii /LINER objects. A pioneering investigation of its ionized gas
velocity field is achieved by Durret \& Warin (1990), in the context
of their study of extended nebulosities surrounding AGNs. The first
deep insight into its very starburst nature comes with the narrow-band
\ha\ imaging survey of Pogge \& Eskridge (1993) who reveal the
presence of a roundish, nuclear star-forming complex, consisting of
bright clumps. Finally, an estimate of the relative contribution of
its very young population component (a few Myr old) has been derived
only recently by Gu et al. (2001), by means of spectral synthesis
techniques.

As far the galaxy's activity is concerned, an optical classification
as a {\em Seyfert~2} nucleus is given by Veilleux et al. (1995) and
such a classification is kept in the 10th edition of the Catalogue of
Quasars and Active Nuclei (Veron \& Veron 2001). A proper description
of the composite (Seyfert$+$\hii ) nature of the object, again on the
basis of optical data alone, can be ultimately found in the work of
Kewley et al. (2001). The full power of the AGN lurking at the center
of NGC~7679 has been finally unveiled in the X-ray domain by means of
both {\em Beppo}SAX and ASCA observations which reveal a bright, and
variable central source (Della Ceca et al. 2001). Unlike previous
classifications, these authors conclude that the only kind of AGN
consistent with its X-ray properties is a {\em Seyfert~1} nucleus.

In the following we make use of narrow-band imaging, optical and UV
spectroscopy, as well as of radio observations to characterize the
role of the plausible close encounter of the two galaxies, to quantify
the central current star formation episode in NGC~7679
and---speculatively---to investigate whether the onset of activity in
both nuclei is related to such past event. In particular, (i) we show
in detail the pattern of star-forming regions close to NGC~7679
nucleus (ii) we compare the inner ionized-gas velocity field of the
galaxy with the extended velocity field of the neutral hydrogen
component (iii) we constrain both age and current star formation rate
of the current starburst and, finally, (iv) try to place this rich
phenomenology in the context of the discussed AGN-starburst
connection.

\section{Observations and data reduction}

\subsection{Optical imaging}

Narrow-band imaging of NGC~7679 was obtained under good seeing
conditions (0.8$''$ FWHM) on September 28, 1999 with the European
Southern Observatory (ESO) Multi-mode Instrument (EMMI) of the 3.58~m
New Technology Telescope (NTT). The detector was the Tektronix TK2048
CCD mounted on the EMMI Red Channel, giving an effective pixel size of
0.27$''$ with a 9$'$.1$\times$8$'$.6 field of view. The on-band and
off-band \ha\ filters were ESO~\#598 and \#596, centered at
$\lambda_c=6685$ \AA\ and $\lambda_c=6568$ \AA\ and having a nominal
FWHM of $\Delta\lambda$~=~67~\AA\ and $\Delta\lambda$~=~73~\AA,
respectively. An exposure time of 600~sec was adopted for both science
frames, while a shorter (200~sec) exposure of the star Feige~110 was
obtained through the on-band filter for calibration purposes.

Using standard MIDAS\footnote{MIDAS is developed and maintained by the
European Southern Observatory.} routines the images were bias
subtracted, corrected for flat field using sky flats and cleaned for
cosmic rays. The sky background level was removed as a constant value
estimated in regions free of sources in the images. The two
consecutive NGC~7679 images obtained through the off-band/on-band
filters, respectively, were shifted and aligned using common field
stars. The point spread function (PSF) of the \ha\ image shows a
slight North-West/South-East elongation not seen in the adjacent
continuum image. After performing several convolution tests to match
the two PSFs, we subtracted directly the scaled off-band continuum
from the on-band image in order to keep as much as possible the fine
detail of the emission-line features allowed by the good seeing
conditions. The best scaling parameter turned out to be very close to
the unity, as expected from the relative efficiencies and bandwidths
for the on-band and off-band filter. A fine tuning of this parameter,
obtained by imposing that the resulting continuum-subtracted image
does not show spurious negative areas, did confirm that the best
choice was to avoid any kind of re-scaling before subtraction.

Flux calibration was derived by means of the observation of the star
Feige~110. Under the assumption that the sensitivity and the absolute
flux for Feige~110 are constant across the on-band filter, following
Sparks et al. (1993) one can estimate the sensitivity function (in erg
cm$^{-2}$ count$^{-1}$) at the redshifted \ha\ on the basis of the
measured star count rate (in count s$^{-1}$) and equivalent width of
the filter. This parameter, in turn, provides the needed conversion to
derive the total emission-line flux F$_{H\alpha+[NII]}$ from the
observed count rate for the galaxy chosen region. The derived flux
has been later corrected for the atmospheric extinction.

\subsection{Optical spectroscopy}
\label{sec:spectroscopy}

The long-slit spectroscopic observations of NGC~7679 were carried out
at the European Southern Observatory in La Silla (Chile) with the ESO
1.52-m telescope (runs~1 and 4-5), at the Mt. Ekar Observatory
in Asiago (Italy) with the 1.82-m telescope (runs~2 and 3), and at the
Multiple Mirror Telescope Observatory in Arizona with the Multiple
Mirror Telescope (MMT, run~6).  The details about the
instrumental setup of each observing run are given below in
Table~\ref{tab:spectroscopy_setup}.

\begin{table*}
\caption{Instrumental setup and log of spectroscopic observations}
\centering
\begin{tabular}{lcccccc}
\hline\hline
\multicolumn{1}{l}{Parameter} &
\multicolumn{1}{c}{Run 1} &
\multicolumn{1}{c}{Run 2} &
\multicolumn{1}{c}{Run 3} &
\multicolumn{1}{c}{Run 4} &
\multicolumn{1}{c}{Run 5} &
\multicolumn{1}{c}{Run 6} \\
\noalign{\smallskip}
\hline
\noalign{\smallskip}
Date                                  & 21 Nov 1987    & 06 Sep 1992      & 25-26 Sep 1994& 26 Sep 1998      & 10 Sep 1999       & 02 Nov 2002     \\
Telescope                             & 1.52-m ESO     & 1.82-m Ekar      & 1.82-m Ekar   & 1.52-m ESO       & 1.52-m ESO        & MMT             \\
Spectrograph                          & B\&C           & B\&C             & B\&C          & B\&C             & B\&C              & Blue Channel    \\
Grating$\rm ^a$ ($\rm gr\;mm^{-1}$)   & \#12 1200      & 1200             & 1200          & \#25 400         & \#33 1200         & 1200            \\
Detector                              & \#13 RCA       & Thomson TH7882   & Thomson TH7882& \#39 Loral-Lesser& \#39 Loral-Lesser & \#22 Loral      \\
Detector size (pixels)                & $1024\times640$& $580\times388$   & $580\times388$& $1024\times1024$ & $1024\times1024$  & $3072\times1024$\\
Pixel size ($\rm \mu m^{2}$)          & $15\times15$   & $23\times23$     & $23\times23$  & $15\times15$     & $15\times15$      & $15\times15$    \\
Pixel binning                         & $1\times1$     & $1\times1$       & $1\times1$    & $1\times1$       & $1\times1$        & $1\times1$      \\
Scale ($\rm ''\;pixel^{-1}$)          & 0.68           & 1.23             & 1.23          & 0.82             & 0.82              & 0.30            \\
Dispersion ($\rm \AA\;pixel^{-1}$)    & 0.864          & 1.025            & 1.025         & 2.81             & 0.985             & 0.49            \\
Slit width ($''$)                     & 1.3            & 3.7              & 5.3           & 2.0              & 2.2               & 1.5             \\
Slit length ($'$)                     & 4.5            & 5.8              & 5.8           & 4.1              & 4.1               & 2.5             \\
Spectral range (\AA)                  & 5840--6730     & 6260--6820       & 6300--6880    & 3350--9300       & 4860--6860        & 4460--5950      \\
Comparison lamp                       & He--Ar         & He--Ar           & He--Ar        & He--Ar           & He--Ar            & He--Ar--Ne      \\
Instr. FWHM (\AA)                     & $1.64\pm0.12$  &$1.73\pm0.04$     & $2.50\pm0.03$ & $9.65\pm0.38$    & $2.99\pm0.01$     & $2.02\pm0.02$   \\
Instr. $\sigma$$\rm ^b$ (\kms)        & 32             & 34               & 49            & 187               & 58                & 53              \\
Seeing FWHM ($''$)                    & 1.0--1.5       & 1.5--2.0         & 1.5--2.5      & 0.8--1.0         & 2.0--3.0          & 1.8--2.2        \\
Observed P.A. ($^\circ$)              & 3,93           & 63,78,93,108,123 & 28,48,138,158 & 0                & 87                & 87              \\
Exposure time (sec)                   & 3600           & 2400             & 3600          & $2\times900$     & $2\times3600$     & 2700            \\
\noalign{\smallskip}
\hline
\noalign{\medskip}
\end{tabular}
\begin{minipage}{17cm}
$\rm ^a$ The grating has been used at the first order.\\
$\rm ^b$ The instrumental velocity dispersion as measured at \ha .\\
\end{minipage}
\label{tab:spectroscopy_setup}
\end{table*}

\begin{figure}
  \centering\includegraphics[width=\columnwidth]{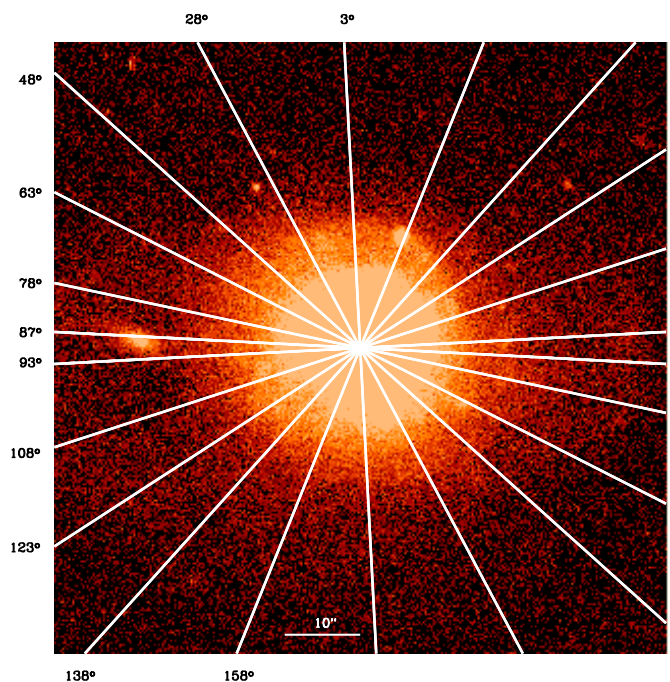}
  \caption{The angular coverage of our medium-resolution
             long-slit spectra of NGC~7679.  The lines mark the
             nominal position of the slits.}
  \label{fig:cover_field}
\end{figure}

Different medium-resolution spectra (runs 1, 2, 3, 5, and 6) were
taken along several axes after centering the galaxy nucleus on the
slit using the guiding camera. Moreover, two low-resolution spectra
(run 4) was taken with the long slit crossing the nucleus along the
North-South direction. An overall picture of the velocity field
sampling assured by our set of medium-resolution spectra is shown in
Fig.~\ref{fig:cover_field}. A lamp spectrum was taken before and/or
after every science exposure for wavelength calibration purposes.

%% FIG.1

Basic data reduction was performed as in Corsini et al. (1999). Using
standard {\tt ESO-MIDAS} routines, all the spectra were bias
subtracted, flat-field corrected by quartz lamp and twilight
exposures, cleaned from cosmic rays, and wavelength calibrated.
The flat-field correction was performed by means of both quartz lamp
and twilight sky spectra in order to correct for pixel-to-pixel
sensitivity variations and large-scale illumination patterns due to
slit vignetting.
Cosmic rays were identified by comparing the photon counts in each
pixel with the local mean and standard deviation and they were
eliminated by interpolating over. Residual cosmic rays were eliminated
by manually editing the spectra.
We checked that the wavelength rebinning was done properly by
referring to the brightest night-sky emission lines in the observed
spectral ranges. The resulting accuracy in the wavelength calibration
is typically 5 \kms.

For each observing run the instrumental resolution was derived in the
spectral region of the \ha\ emission line as the mean of Gaussian
FWHMs measured for a number of unblended arc-lamp lines of a
wavelength-calibrated comparison spectrum. The mean FWHM of the
arc-lamp lines and the corresponding resolution at \ha\ are given in
Table~\ref{tab:spectroscopy_setup}.
After the calibration, all the spectra were corrected for CCD
misalignment.  The contribution of the sky was determined from the
outermost $\sim20''$ at the two edges of the resulting frames where
the galaxy light was negligible, and then subtracted.  In runs 4
and 5 the two spectra obtained at $\rm PA=87^\circ$ were coadded
using the center of the stellar-continuum radial profile as reference.
In run 4 we observed some spectrophotometric standard stars to allow
the flux calibration of the low-resolution spectra.

\subsection{Ultraviolet spectroscopy}
\label{sec:uv_spectroscopy}

The occurrence of a powerful star formation event in NGC~7679 is made
clear by the availability in the International Ultraviolet Explorer
(IUE) archives of a long-exposure ($\sim$7h), large aperture
($\sim$10$''$$\times$20$''$) SWP spectrum covering the
wavelength range 1200-1900~\AA. Unlike the majority of IUE spectra of
nearby galaxies, the relatively-high redshift of this object allows
the Ly$\alpha$ emission of its disk to appear fully detached from the
contaminating geocoronal Ly$\alpha$. Such intrinsic line is indeed
quite strong and extended ($\sim$12$''$) along the spatial direction
(at P.A. = 134$^\circ$).

The IUE spectrum has been re-extracted by means of our own MIDAS
routines. We started from the so-called NEWSIPS spatially-resolved,
low-resolution image (SILO), de-archived from the Italian National
Host of INES (IUE Newly Extracted Spectra) distribution system. The
properly fluxed, redshift and galactic reddening corrected UV spectrum
is presented and discussed below.  The adopted foreground galactic
extinction was E(B$-$V)=0.06 from Burstein \& Heiles (1984).

\subsection{Radio observations}
\label{sec:radio}

The presence of an extended \hi\ halo around NGC~7679 {\em and} its
companion galaxy NGC~7682 has been pointed out by observations carried
out at Arecibo Observatory since 1986 (Duprie \& Schneider 1996). They
derive for either galaxy a neutral gas mass
M$_{HI}$$\sim$5$\times$10$^9$ M$_\odot$.  A subsequent work of
Kandalyan (2003), taking into account also radio CO line observations,
indicates that NGC~7679 does possess also a comparable amount of
molecular hydrogen (M$_{H2}$$\sim$6$\times$10$^9$ M$_\odot$).

Our recent observations of NGC~7679 and its neighbor NGC~7682 in the
21-cm line of \hi\ were obtained on July 28 and August 20, 2000 using
the D configuration ({\em i.e.} the best-suited to detect faint
extended emission) of the NRAO Very Large Array (VLA)\footnote{The
National Radio Astronomy Observatory is operated by the Associated
Universities, Inc., under cooperative agreement with the National
Science Foundation.}. Both galaxies, with an angular separation of
$\sim7\farcm5$, easily fit into the field of view of the VLA, which
has half-power beamwidth at 21 cm of $\sim1'$. The VLA correlator was
configured so as to provide a velocity resolution of $\sim$21~km
s$^{-1}$.  The total on-source time was $\sim$2.0 hours. The data were
calibrated, continuum subtracted, and mapped in the standard fashion
using the Astronomical Image Processing System (AIPS).

\section{Results}

\subsection{The optical morphology}

\begin{figure}
\centering\includegraphics[width=\columnwidth]{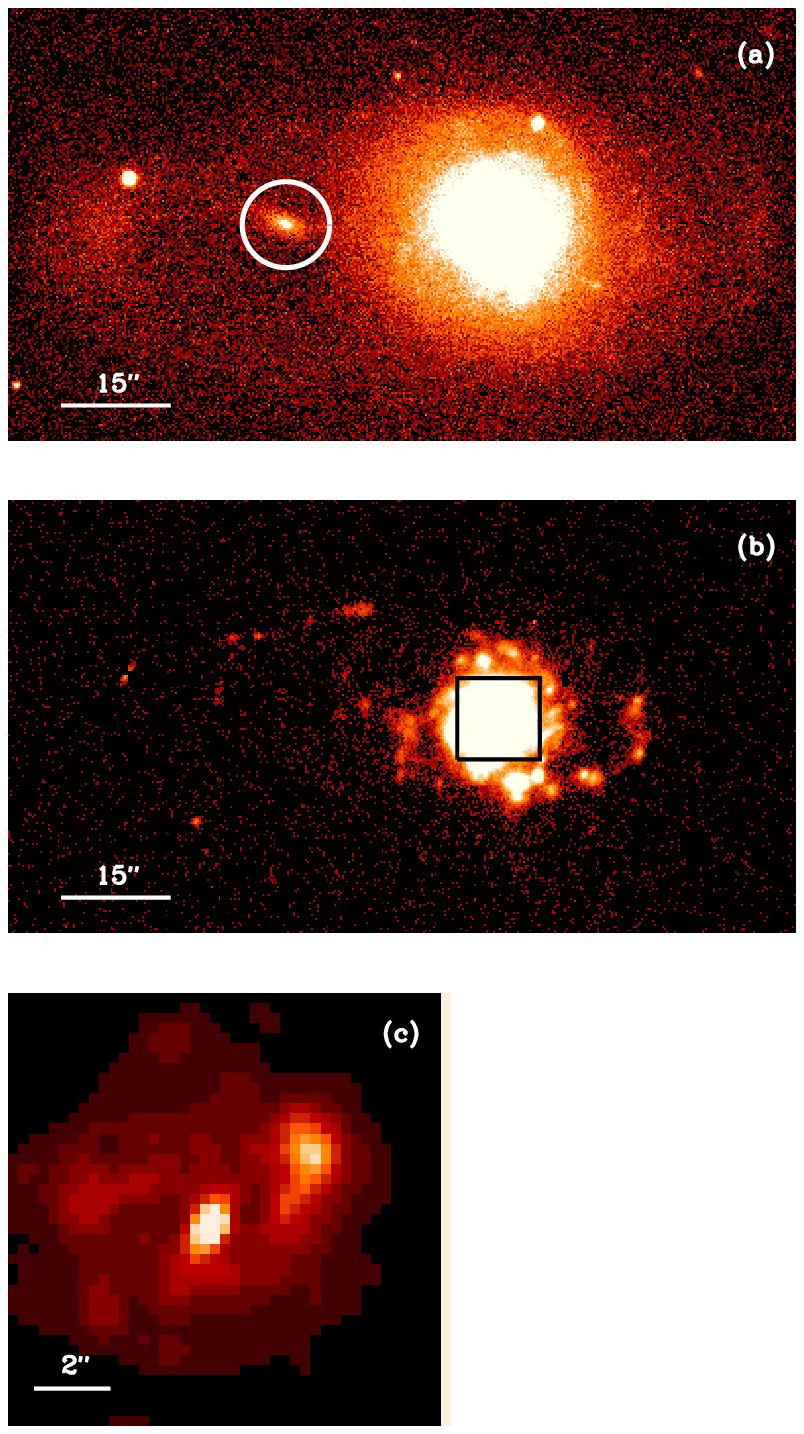}
  \caption{{\em (a):} Narrow-band continuum image of NGC~7679 and
            its surrounding distorted arm. North is up and East on the
            left. The circled object is a background galaxy not
            involved in the outer tidal distortion of NGC~7679 (see
            discussion below). {\em
            (b):} Pure \ha$+$\nii\ image of the same field showing
            individual bright knots and loose gaseous debris; the
            squared inset is zoomed in panel {\em (c)} in order to
            show the circumnuclear emitting region.}
  \label{fig:imaging}
\end{figure}

The NGC~7679 continuum, the overall continuum-subtracted pure emission
image, and innermost spiral-like circumnuclear emitting region are
shown in Fig.~\ref{fig:imaging}.

%% FIG.2

The distribution of the ``pure'' \ha$+$\nii\ emission, consists mainly
of a roundish component hiding a brighter, very central spiral-like
structure. Thanks to the higher NTT/EMMI dynamical range and resolution,
such an image represents an evident improvement over the emission-line
image recorded by Pogge \& Eskridge (1993).

In particular the circumnuclear complex morphology, indicates that the
starburst activity goes down to the very central regions. Moreover,
both its appearance and absolute size are reminiscent of the
circumnuclear starburst in NGC~5248 that is interpreted as induced by
a stellar bar in the very central region as recently discussed by
Jogee et al. (2002).

\subsection{The ionized-gas velocity field}

\begin{figure*}
  \centering
  \includegraphics[width=8.5cm]{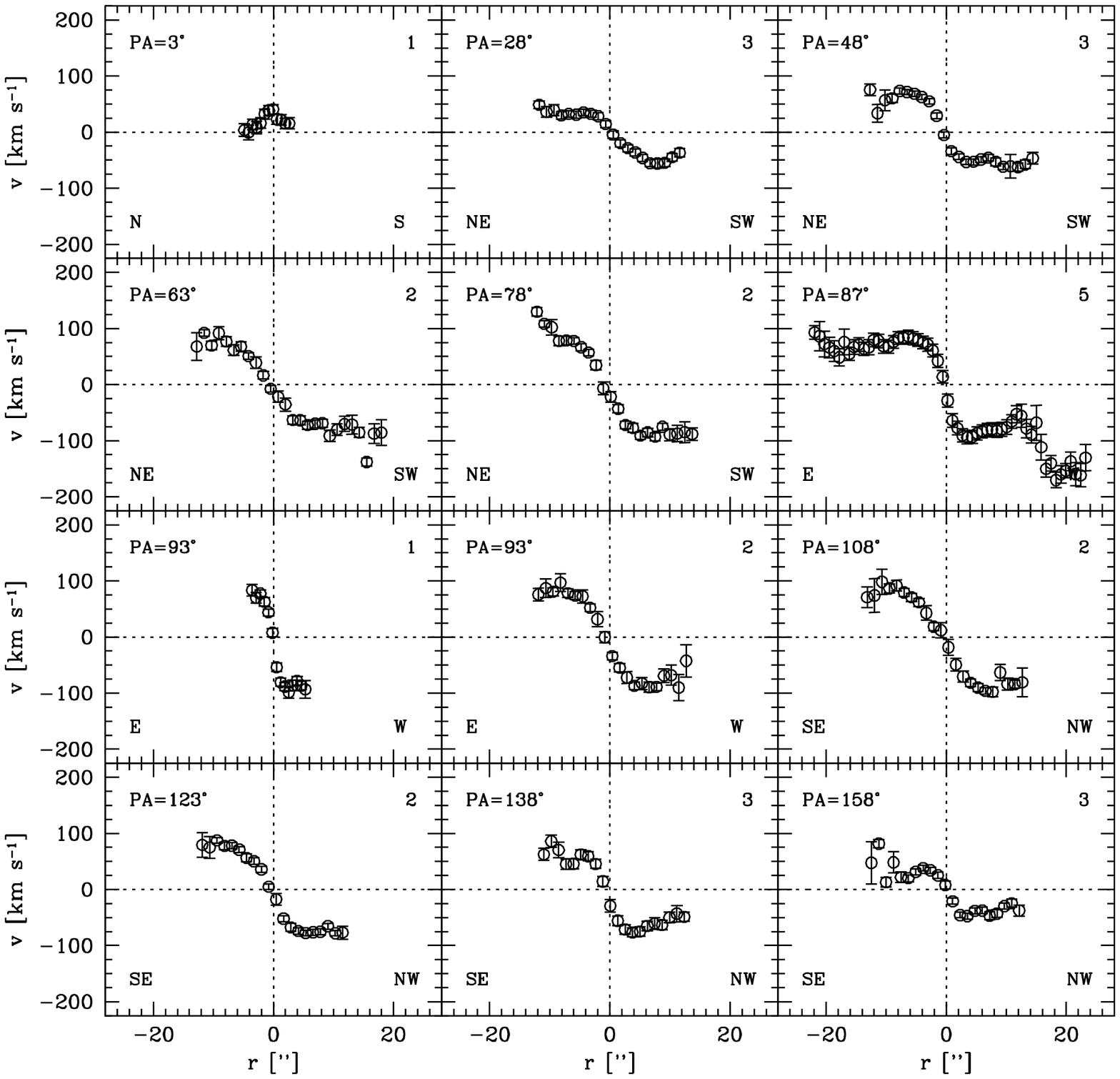}
  \includegraphics[width=8.5cm]{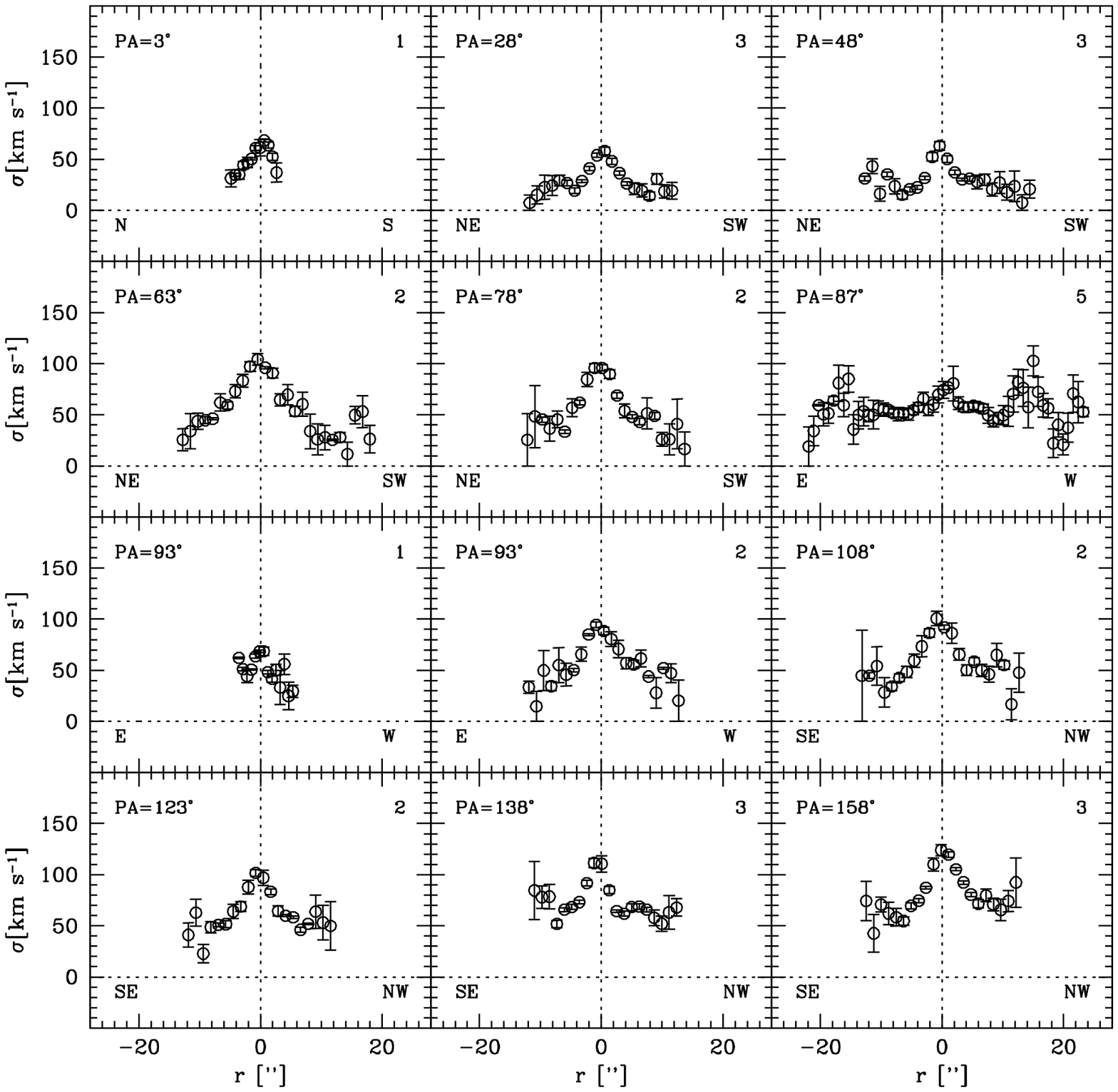}
   \caption{Ionized-gas rotation velocity curves ({\it left panels\/})
            and velocity dispersion profiles ({\it right panels\/})
            measured along the observed axes of NGC 7679. Plotted
            velocities are subtracted by systemic velocity and not
            corrected for galaxy inclination.  Each subframe contains
            both the corresponding P.A. ({\it upper left\/}) as well
            as the observing run number ({\it upper right\/}; see
            Tab.~1). }
\label{fig:kinematics}
\end{figure*}

The rotation curves and the velocity dispersion profiles
we measured along the observed axes of NGC~7679 are shown in
Fig.~\ref{fig:kinematics}.

%% FIG.3

The line-of-sight velocities and velocity dispersions of the
ionized-gas component were measured by means of the {\tt MIDAS}
package {\tt ALICE}.
At each radius, we measured the \ha , \niipg\ and \siipg\ emission
lines where they were clearly detected.  The position, the FWHM, and
the uncalibrated flux of each emission line were determined by
interactively fitting one Gaussian to each line plus a polynomial to
its local continuum.
The central wavelength of the fitting Gaussian was converted into
velocity in the optical convention and the standard heliocentric
correction was applied.

The heliocentric systemic velocity ($V_\odot=5165\pm5$ km s$^{-1}$)
was derived from the center of symmetry of the rotation curve along
the galaxy major axis. It corresponds to $V_{3K} = 4796$ km s$^{-1}$
after applying the 3K background correction following Fixsen et
al. (1996).  Note that the latter value derived by us is consistent,
within the error, with the previous, analogously corrected RC3
velocity given in the introduction.
The Gaussian FWHM was corrected for the instrumental FWHM, and then
converted into the velocity dispersion. In the regions where the
intensity of the emission lines was low, we binned adjacent spectral
rows in order to improve the signal-to-noise ratio of the lines.

The velocity profiles presented in Fig.~\ref{fig:kinematics} display a
high degree of simmetry within the central regions, suggesting that
they may be well represented by a simple disk geometry. To determine
whether a regular disk rotation can explain the observed gas
kinematics we used the {\em kinemetry} software by Krajnovi\'c et
al. (2005) to construct the best fitting non-parametric disk model of
the gas kinematics within the innermost 15$''$ from the galaxy
nucleus. The idea behind our procedure is that for a disk that is
observed at an inclination $i$, an ellipse of axial ratio $q=\cos i$,
and the same position angle as the disk, will sample equal radii in
the disk plane. Moreover, if the motion is purely circular, the
observed velocity along that ellipse will be described by a cosine
law. Our procedure consists of iteratively changing the inclination
until these two conditions are satisfied.  In more detail we performed
the following steps:

\begin{enumerate}
\item We assumed an observed inclination $i$ and a projected
position angle (PA) for the gas disk;

\item We sampled the observed velocity $V$ along a set of ellipses,
regularly spaced in the semimajor axis, with constant $q=\cos i$ and
PA as defined above, using linear interpolation when needed to
estimate the velocity;

\item Along every ellipse of semimajor axis $R$
we least-squares fitted the velocity with the formula $V_{\rm
d}(R,\theta)=V_0(R)\cos\theta$, which describes the line-of-sight
velocity along the disk where the circular velocity is $V_{\rm
c}(R)=V_0(R)/\sin i$. Here $\theta$ is the eccentric anomaly, measured
from the projected major axis of the ellipse;

\item We computed the model velocity $V_{\rm d}(x,y)$
at the observed $N$ coordinates $(x,y)$ on the sky using linear
interpolation;

\item We determined the agreement between the
observed velocity field and the disk model as $\chi^2=\sum_{j=1}^{N}
[(V_{{\rm d},j}-V_j)/\Delta V]^2$, where $\Delta V$ are the
measurements errors;

\item We iterated the points 1--5 for a
regular grid of $q$ and PA values. We finally selected the best
fitting values and the corresponding errors from the contours of
$\chi^2(q,{\rm PA})$.
\end{enumerate}

\begin{figure}
\centering\includegraphics[width=0.8\columnwidth]{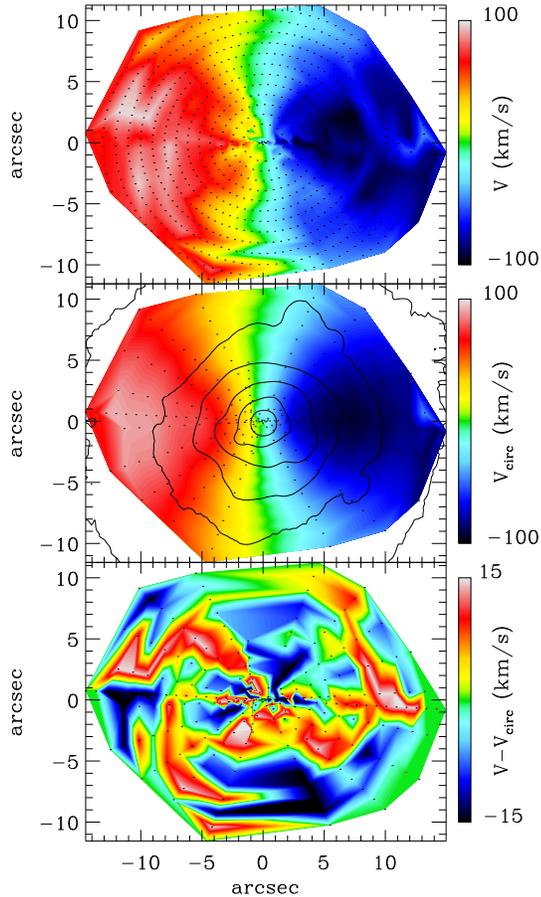}
  \caption{Interpolated velocity field and sampling
             ellipses {\em (top panel)}, best fitting velocity field
             for a disk model {\em (middle panel)}, and residuals
             between interpolated and best fitting velocity
             field {\em (bottom panel)} for the ionized gas component
             of NGC~7679 within $15''$ from the center.}
 \label{fig:kinemetry}
\end{figure}

The result of the above procedure provided the best fitting disk
parameter $q=\cos(i)=0.90\pm0.05$ and ${\rm PA}=93\pm3$ (3$\sigma$
errors). The interpolated data and the ellipses along which the
velocity was sampled are presented in the top panel of Fig.~\ref{fig:kinemetry},
while the best fitting disk model is shown in the middle panel. The
residuals between the data and the best fitting model are displayed in
the bottom panel. It is apparent that the observation can be
remarkably well described by the adopted simple thin disk
approximation. The best fitting $\chi^2\sim2$ also indicates that the
gas kinematics in the central is marginally consistent with a pure
nearly face-on ($i\approx26^\circ$) disk model. No obvious regular
features are apparent in the residual map, suggesting that the
difference of the $\chi^2$ from unity is due only to residual
systematics in the different data sets. Unfortunately the near face-on
geometry of the observed disk, and the small spatial extension of the
da ta, is unfavorable for a detailed mass modeling of this galaxy, due
to large uncertainties in the mass deprojection and inclination
effects. We will not explore this further in this paper.

%% FIG.4

\subsection{The background false interloper}

A particularly intriguing feature of NGC~7679 morphology is that the
most disturbed side of the galaxy's body coincides with the position
($\sim$29$''$ East) of a much fainter, slightly elongated companion
one could easily misinterpret as an interacting satellite galaxy
capable of distorting the outer regions of the main galaxy. The
projection effect is so convincing as to induce Pogge \& Eskridge
(1993) to ascribe to the interaction with this object the most
pronounced tidal structures of NGC~7679. The same oversight can be
noticed also in the later study by Della Ceca et al. (2001).

Thanks to a long-slit spectrum obtained in run 6 with MMT we can now
unmistakably demonstrate that such an ``intruder'' is actually a
background galaxy with a measured heliocentric recession velocity
$V_{\odot} = 33105\pm5$ km s$^{-1}$ which corresponds to $V_{3K} =
32736$ km s$^{-1}$. after applying the 3K background correction
following Fixsen et al. (1996). By adopting the proper relativistic
redshift formula we derive z=0.1158, {\em i.e.} approximately
7$\times$ the observed redshift of NGC~7679.

\begin{figure}
\centering\includegraphics[width=\columnwidth]{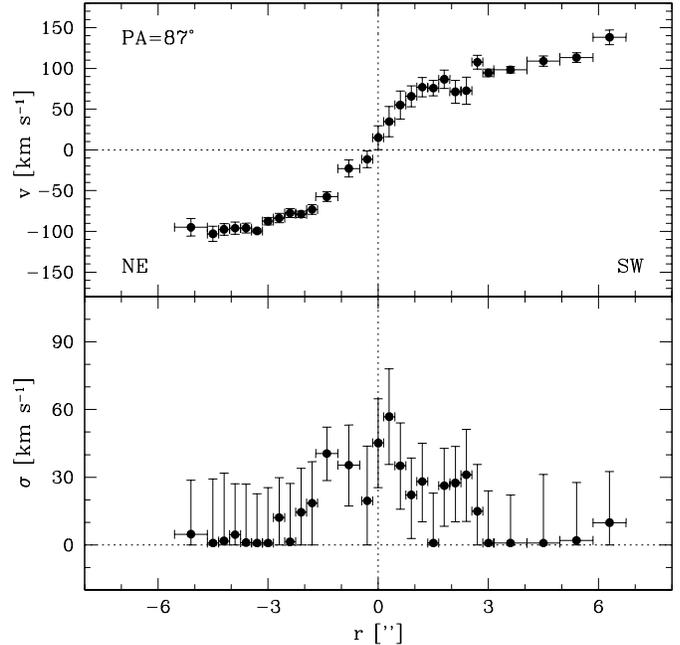}
  \caption{The derived gaseous rotation curve and velocity
             dispersion profile of the background object projected on
             the outskirts of NGC~7679. The scale is $\sim2$~kpc
             arcsec$^{-1}$.}
  \label{fig:rot_intrud}
\end{figure}

Assuming $\Omega_M=0.3$ and $\Omega_\Lambda=0.7$ ({\em cf.} Tegmark et
al. 2004), we derive an angular diameter distance of
$\sim$400~Mpc. Since the corresponding angular scale is $\sim$1.94~kpc
arcsec$^{-1}$, its observed ionized-gas rotation curve extends out to
R$\sim$11.6~kpc (see Fig.~\ref{fig:rot_intrud}). The mass within such
a radius turns out to be $\sim3\times10^{10}$ M$_\odot$ assuming a
spherical mass distribution and the galaxy seen edge-on.

%% Fig.5

\subsection{The radio morphology}

As pointed out above, NGC~7679 has previously been detected in a
number of single-dish observations (see Duprie \& Schneider 1996 and
references therein). Duprie \& Schneider (1996) reported that the \hi\
line profile of the galaxy is peculiar, but cautioned against possible
contamination from the neighbor Seyfert galaxy NGC~7682.

\begin{figure*}
  \centering
  \includegraphics[width=17cm]{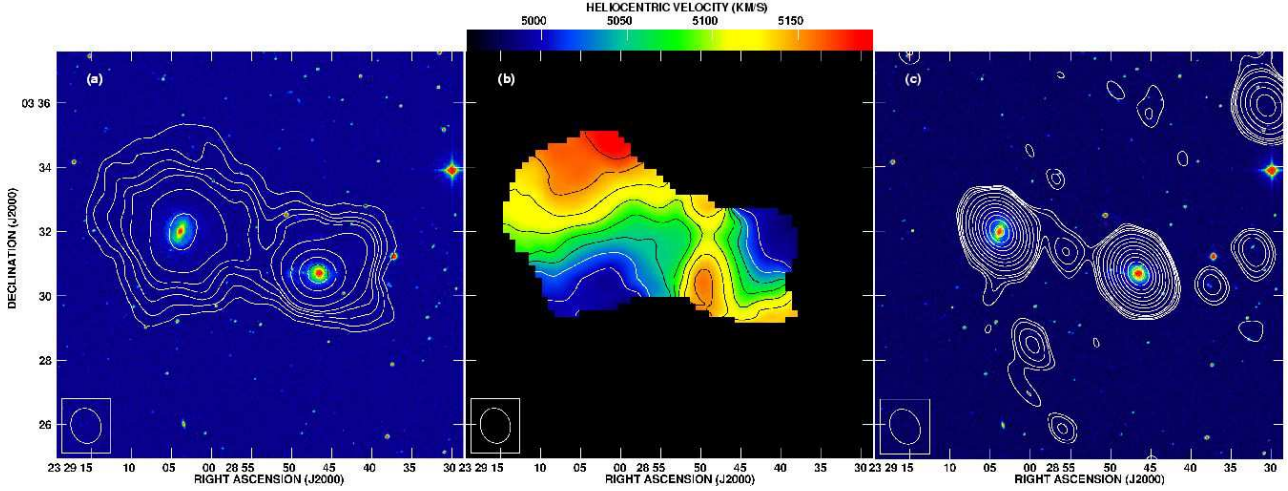}
  \caption{{\em (a)}: \hi\ total intensity (moment 0) map superposed
            to the optical image of NGC~7679 (lower right) and
            NGC~7682 (upper left). The peak flux is 2.4 $\times$
            10$^{3}$ Jy beam$^{-1}$ m s$^{-1}$. Contour levels are
            (0.3, 0.9, 1.5, 2.1, 2.7, 4.5, 9, 15, 21) $\times$
            10$^{2}$ Jy beam$^{-1}$ m s$^{-1}$.
            {\em (b)}: \hi\
            velocity field (moment 1) contour map on the total
            intensity (moment 0) color-coded map. Coded colors (from
            blue to red) are for 4750, 4775, 4800, 4825, 4850, 4875,
            4900, 4925, 4950, 4975, 5000, 5025 km s$^{-1}$.
            {\em (c):} \hi\ continuum emission map. The peak flux
            is 5.3 $\times$ 10$^{-2}$ Jy beam$^{-1}$ m
            s$^{-1}$. Contour levels are (3.6, 4.8, 7.2, 9.6, 19.2,
            36, 60, 96, 144, 216, 288, 384, 480) $\times10^{-4}$ Jy
            beam$^{-1}$ m s$^{-1}$.
            The synthesized beam of all the maps is
            $68\farcs0\times53\farcs3$.   }
\label{fig:hi_full}
\end{figure*}

Our \hi\ image (Fig.~\ref{fig:hi_full}{\em (a)}) shows that NGC~7679
and NGC~7682 share a common gaseous envelope: an examination of the
channel maps (Kuo et al., in preparation) demonstrates that this
envelope comprises a tidal bridge connecting the two galaxies, as well
as opposing tidal tails from the two galaxies. The extended and
diffuse optical feature on the eastern side of NGC~7679
(Fig.~\ref{fig:kinematics}{\em (a)}) can be identified with the inner
region of this \hi\ tidal bridge.

%%% FIG.6

The total flux density of each of the three components are
$54.6\pm0.9$ mJy for NGC~7679, $52.6\pm1.0$ mJy for NGC~7682,
$1.1\pm1.4$ mJy for the bridge. Assuming our adopted distance of 63.9
Mpc, their radio continuum luminosities are $2.67\times10^{29}$ erg s$^{-1}$ Hz,
$2.58\times10^{29}$ erg s$^{-1}$ Hz, $4.95\times10^{27}$ erg s$^{-1}$
Hz, respectively. The continuum emission associated with each Seyfert
galaxy is not spatially resolved as seen in
Fig.~\ref{fig:hi_full}{\em (c)}.
The latter shows an image of the continuum
emission at 21 cm, most probably produced or dominated by synchrotron
emission from relativistic electrons.  The two strongest sources
visible are those associated with NGC~7679 and NGC~7682.
Interestingly, a continuum source is visible lying between NGC 7679
and 7682, and perhaps forming a bridge between these two objects.
This source or bridge coincides with the \hi\ tidal bridge connecting
the two galaxies, and may be produced by star-formation activity in
this bridge.  Of course, we cannot rule out the possibility that this
source is an unrelated foreground or background source.

\subsection{The cold gas velocity field}

Although NGC~7682 does not appear to be disturbed in the optical, not
only it does exhibit extended \hi\ tidal features, but its \hi\
kinematic major axis shows a wide misalignment ($>$50$^\circ$) with
its optical major axis (see Fig.~\ref{fig:hi_full}{\em (b)}).  The
highly extended tidal features indicate that this system must have
started to gravitationally interact some time ago, and can be used to
estimate the timescale since closest approach. More precisely, tidal
features seen in both galaxies extend to about 80~kpc from each
galaxy, and have a measured maximum radial velocity of $\sim$120 km
s$^{-1}$.  If the latter also is their maximum velocity in the plane
of the sky, and assuming that the gas originates from the outer
regions of a disk with radius of $\sim$20 kpc, the tidal structures
would therefore have a kinematic age of $\sim$500 Myr.

\subsection{Ultraviolet energy distribution and burst dating}

In Seyfert~2 galaxies the optical energy distribution toward shorter
wavelengths is increasingly dominated by the so-called ``featureless
continuum'' (FC; {\em cf.} Heckman et al. 1995). A fraction of it has to be
ascribed to the reflected (polarized) light by a ``hidden'' Seyfert~1
nucleus (FC1) while the majority of it (FC2; appearing as a blue,
unpolarized continuum) is currently thought to be originated by a
population of hot, massive stars. Understanding the relative role of these
two components is quite important for a system like NGC~7679, owing to
its highly composite nature.

\begin{figure}
\centering\includegraphics[width=\columnwidth]{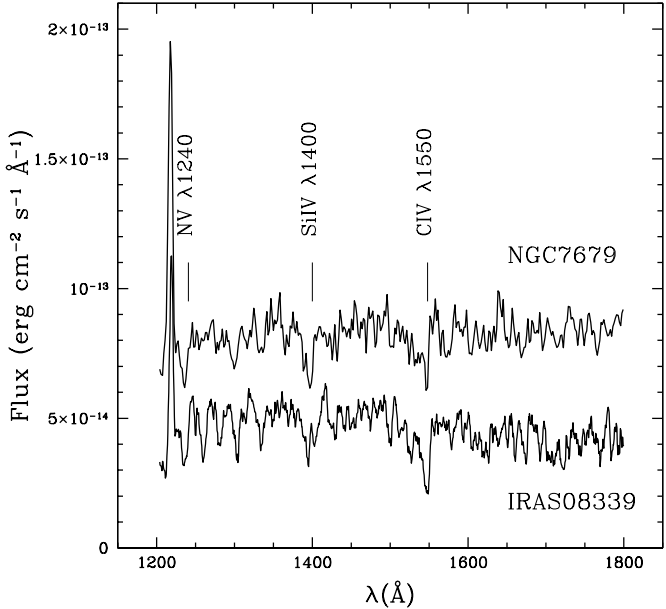}
  \caption{Comparison of the UV spectra of NGC~7679 (IUE) and the
             young starburst IRAS~0339$+$6517 (HUT) in the region
             where they overlap ($\lambda\lambda$~1200-1800\AA). Both
             spectra have been corrected for redshift and galactic
             extinction as given by Burstein \& Heiles (1984).  The
             HUT spectrum of IRAS~08339$+$6517 has been smoothed with
             a 7~pixel window. A constant amount of
             4$\times$10$^{-14}$ flux units has been added to the
             NGC~7679 UV energy distribution---previously normalized
             to match the HUT spectrum around $\lambda$~1500~\AA---for
             showing purposes. The most prominent stellar wind
             absorption lines are marked.}
\label{fig:iras}
\end{figure}

In order to remove the degeneracy of the two above continua observed
at optical wavelengths for the youngest populations ($<$10~Myr; {\em
cf.} Storchi-Bergmann et al. 2000), one has to move to the UV region
where an ongoing starburst, when present, reveals unambiguously
several strong absorption stellar wind resonance lines, such as
N~{\small V}~$\lambda$1240, Si~{\small IV}~$\lambda$1400 and
C~{\small IV}~$\lambda$1550, overposed to the underlying continuum. In
this respect, the continuum originated by very young stars (FC2)
should not be named ``featureless'' at all.  Besides showing the above
strong stellar-wind lines, the NGC~7679 IUE spectrum
(Fig.~\ref{fig:iras}) reveals that C~{\small IV}~$\lambda$1550 does
possess an outstanding P-Cygni profile (instead of being pure
photospheric absorption), thus assuring that the burst we are
observing is still within its early phase, dominated by strong stellar
winds. Within the IUE large aperture (10$''$$\times$20$''$), such a
burst gives origin to a extinction-corrected Ly$\alpha$ flux
F$_{Ly\alpha}$ = 4.9 $\times$ 10$^{-13}$ erg cm$^{-2}$ s$^{-1}$, whose
equivalent width is 37~\AA.

%% FIG.7

A reliable estimate of the age of the recently formed stars (and thus
an estimate of the epoch of the starburst onset in NGC~7679) can be
derived by comparing its IUE/SWP spectrum with the overlapping portion
of Hopkins Ultraviolet Telescope (HUT) spectra of similar starburst
galaxies presented by Leitherer et al. (2002). In particular, one can
easily notice the good match of NGC~7679 with both the continuum
and---allowing for more pronounced P-Cygni effects---the UV absorption
features of their young starburst IRAS~08339$+$6517 (see
Fig.~\ref{fig:iras}).

The match between the UV energy distributions of the two
starbursting objects presented in Fig.~\ref{fig:iras} is quite
impressive. Both spectra have been corrected for redshift, galactic
and internal extinction. The latter value has been assumed to be close
to E(B$-$V)$\sim$0.3 for both galaxies. In fact in the case of
IRAS~08339$+$6517 the absorption A$_{1500}$=2.34~mag estimated by
Leitherer et al. (2002) corresponds to E(B$-$V)$\sim$0.29, if one
applies the extinction curve of Savage
\& Mathis (1979), while for the overall line-emitting disk of NGC~7679 one
can assume a value E(B$-$V)$\sim$0.33 obtained by the observed ratio
Ly$\alpha$/\ha $\sim$0.9 to be compared with the theoretical ``Case
B'' ratio Ly$\alpha$/\hb$=31.6$ for T$_{e}$=10$^4$ K and
$\log{n_e}=4$, having assumed the standard Balmer decrement \ha/\hb =
2.86 (see Dopita \& Sutherland 2003).
This strongly suggests that we are
comparing young-star dominated systems very close in age.
In particular, Leitherer et al. (2002) ascribe the UV spectrum of
IRAS~08339$+$6517 to a 10~Myr-old stellar population of solar
composition, though compatible with an age between 5 and 20~Myr. As a
consequence, we feel allowed to date back to 5-20~Myr ago also the
onset of star-formation episode we are witnessing today in NGC~7679.
For comparison, the internal extinction we derive within the slice
($2\farcs0\times2\farcs4$ at $\rm PA=0^\circ$) of the NGC~7679 covered
by our own low-resoultion spectrum (run 4) by measuring its Balmer
decrement, turns out to be higher (E[B$-$V]$\sim$0.60).  Given the
tiny portion of the emitting region recorded by this latter spectrum,
likely affected by local, innermost dishomogeneities, we keep
E(B$-$V)$\sim$0.33 as the representative absorption of the NGC~7679
disk as a whole.

\subsection{The dominance of the starburst at IR wavelengths}

The existence of an ongoing starburst in a galaxy reflects into an
enhancement of the far-infrared (FIR) continuum emission, due to dust
re-radiation of ultraviolet photons from hot OB stars. At the same
time, in the case of a composite object like NGC~7679, one has to
expect a substantial contribution to the FIR continuum by the
underlying AGN. In order to identify which source does dominate the IR
emission in our object, one can resort to the classification scheme
adopted by Mouri \& Taniguchi (2002) on the basis of far-infrared flux
densities of the IRAS database.

Having defined F$_{IR}$ the observed flux between 40 and 120 $\mu$m
and F$_B$ the monochromatic flux at 4300~\AA , respectively, the IR
data of Kewley et al. (2001) imply for NGC~7679 a ratio
F$_{IR}$/F$_B$$>$4. This, in turn, implies that our object belongs to
the Mouri \& Taniguchi (2002) class of {\em starburst-dominated}
Seyfert galaxies. For comparison, the nearby Seyfert~2 galaxy
NGC~7682, likely interacting with NGC~7679, has to be classified
within the above scheme as an {\em AGN-dominated} Seyfert, owing to
its measured F$_{IR}$/F$_B$$<$1 (Mouri \& Taniguchi 2002). As such,
the two systems, though sharing the nuclear Seyfert phenomenon, appear
quite different about the mutual dominance of the central AGN and the
young star emission.

\subsection{Star formation rate}

Correcting for the galactic extinction by Burstein \& Heiles (1984),
our pure emission-line image of NGC~7679 gives a extinction corrected
total flux F$_{H\alpha+[NII]}$ = 8.0 $\times$ 10$^{-13}$ erg cm$^{-2}$
s$^{-1}$. The innermost ($r\leq3''$) region ($r\leq3''$)---{\em i.e.}
the central spiral pattern---gives a corresponding flux
F$_{H\alpha+[NII]}$ = 5.9 $\times$ 10$^{-13}$ erg cm$^{-2}$
s$^{-1}$. As such, the latter region alone provides approximately 75\%
of the total \ha$+$\nii\ emitted flux.
The spectrophotometry of Contini et al. (1998), showing that the
fraction of light due to \ha\ amounts to 64\% of the total
\ha$+$\nii\ emission, allows us to approximately correct such flux
for the contamination of the \niipg\ doublet emission, thus giving a
``pure'' \ha\ flux received from NGC~7679 of 5.1 $\times$ 10$^{-13}$
erg cm$^{-2}$ s$^{-1}$.
Moreover, one has to assume that the latter value is still affected by
some amount of internal extinction, which, in localized regions of
starburst galaxies, can be as high as A(\ha )$\sim$2 mag ({\em cf.}
Kennicutt 1998). In the specific case of NGC~7679, assuming
E(B$-$V)=0.33 as the color excess due to its internal extinction close
to the center (see \S 3.6), the intrinsic \ha\ flux turns to be 1.1
$\times$ 10$^{-12}$ erg cm$^{-2}$ s$^{-1}$. The latter value, properly
transformed into the corresponding L$_{H\alpha}$, allows, in turn, to
estimate the current star formation rate.

On the basis of our adopted distance of 63.9~Mpc, the above flux
translates into a \ha\ luminosity of 5.4$\times$10$^{41}$ erg
s$^{-1}$. By assuming the calibration of Kennicutt et al. (1994) this
implies a SFR $\sim$4 M$_\odot$ yr$^{-1}$. An independent SFR estimate
can be derived also by the UV continuum measured in the IUE
spectrum. After correcting for redshift, galactic and extinction, the
measured average flux in the absorption-free wavelength range between
1425 and 1515~\AA\ is F$_\lambda$= 2.6$\times$10$^{-13}$ erg cm$^{-2}$
s$^{-1}$ \AA$^{-1}$. This translates into a luminosity of
9.6$\times$10$^{28}$ erg s$^{-1}$ Hz$^{-1}$ corresponding, with the
calibration given by Kennicutt (1998), to a SFR $\sim$13 M$_\odot$
yr$^{-1}$. The two levels of star formation ({\em i.e.} that one
derived from the optical recombination line and that obtained by the
UV continuum) are then quite similar and appear relatively high, being
comparable to that shown by other similarly disturbed starburst
systems like, for instance NGC~7673 (SFR$\sim$10-20 M$_\odot$
yr$^{-1}$; Homeier et al. 2002).

As far as radio data are concerned, using the empirical correlation
between the radio continuum luminosity at 1.4 GHz and SFR (Condon
1992; Lou \& Bian 2005) we find a SFR$\sim$7 M$_\odot$ yr$^{-1}$ for
NGC~7679, 6 M$_\odot$ yr$^{-1}$ for NGC~7682, and 0.1 M$_\odot$
yr$^{-1}$ for the bridge region. For both NGC~7679 and NGC~7682, the
computed SFR are, of course, upper limits as a part the radio emission
could from the central AGN. For NGC 7679 such an estimate is within
the range derived from optical and UV measurements.

\subsection{The close encounter with NGC~7682}

Having definitely excluded any role of the close {\em background}
companion in the origin of distorted morphology of NGC~7679, the most
likely culprit becomes NGC~7682, the face-on barred spiral located
only $\sim4\farcm5$ North-East. This latter galaxy---itself a
Seyfert~2 system with an almost identical redshift---offers
straightforward evidence, thanks to the radio observations discussed
above, of a tidal interaction happened a few hundreds million years
ago. Though interaction is often invoked as a starburst-triggering
phenomenon ({\em cf.} Greene et al. 2004), whether this specific
encounter played a role in the onset of current central starburst in
NGC~7679 remains difficult to say, owing the evident youth of the
latter phenomenon.

\section{NGC~7679 in the framework of the AGN-starburst
 connection issue}

NGC~7679, showing simultaneously a vigorous starburst in the optical
and a powerful AGN in the X-ray waveband, is a prototypical composite
object.

\begin{figure}
\centering\includegraphics[width=\columnwidth]{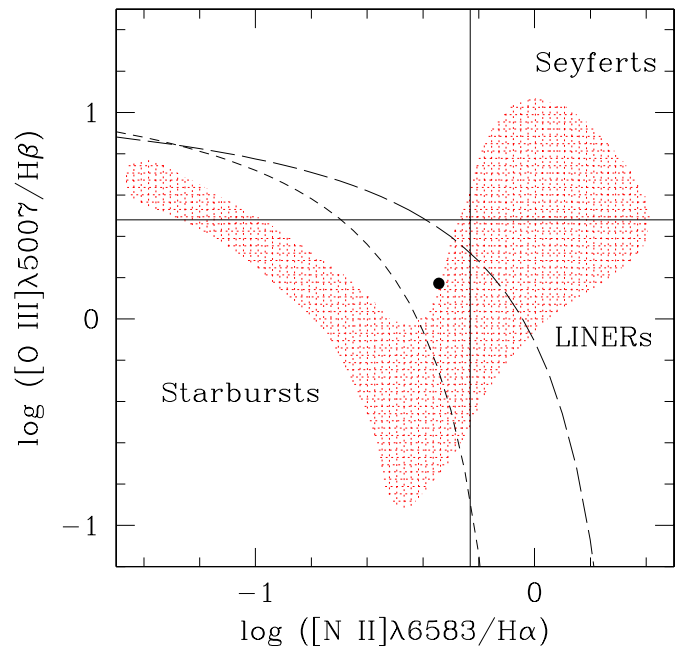}
  \caption[]{The location of NGC~7679 in the optical emission-line
         diagnostic diagram of Kauffmann et al. (2003; filled circle).
          The short
           and long dashed lines mark the demarcation between
           starburst galaxies and AGNs according to Kauffmann et
           al. (2003) and Kewley et al. (2001), respectively. The
           vast majority of the galaxies studied by Kauffmann et
           al. (2003) are located in the hatched region.}
  \label{fig:kauffmann}
\end{figure}

We derived extinction-corrected diagnostic ratios from optical
emission lines, namely log(\oiii /\hb )=0.172 and log(\niig/\ha )
=-0.343 . Such observed ratios are widely consistent with the earlier
corresponding values measured by Dahari (1985). On the basis of the
most recent diagnostic diagrams by Kauffmann et al. (2003) and its
very low \oiii\ ($L_{\rm [OIII]} = 2.12\times10^5$ L$_\odot$) {\em
i.e.} the main parameter they adopt as AGN activity tracer, NGC~7679
falls within closer the region of proper starburst galaxies
(Fig.~\ref{fig:kauffmann}) also in agreement with previous results by
Panessa \& Bassani (2002).

%% FIG.8

\begin{figure}
\centering\includegraphics[width=\columnwidth]{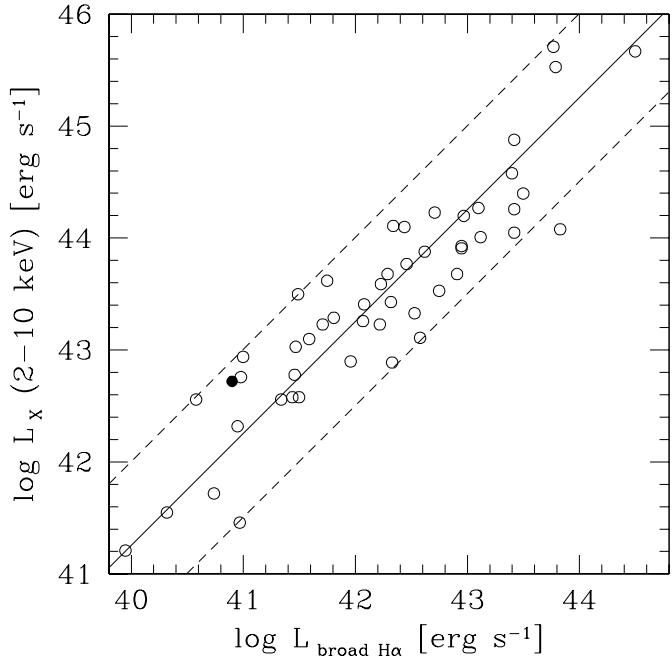}
  \caption{The location of NGC~7679
            in the X-ray/optical diagnostic diagram of Imanishi \&
            Terashima (2004; filled circle). Luminosities of NGC~7679
            are taken from Della Ceca et al. (2001).  The region of
            dust-unabsorbed Type~1 AGNs (open circles) is marked with
            dashed lines.}
  \label{fig:imanishi}
\end{figure}

Given the dominance of star formation at optical wavelengths for
composite systems, specific indicators capable of tracing ``pure'' AGN
power by disentangling it from superposed vigorous star formation are
required. Two useful indicators are currently known, namely the
absorption-corrected intrinsic 2-10~keV X-ray luminosity and the
luminosity of broad hydrogen emission lines which come from the
high-velocity gas around the central supermassive black hole in the
AGN. A recent study of the correlation between these two quantities
(for type-1 AGNs) is given by Imanishi \& Terashima (2004).  Although
the quality of our own optical spectra of NGC~7679 are not suitable to
study in detail its faint broad Balmer component, we resorted---for
its classification---to the high $S/N$ spectrum of Kewley et
al. (2001), who identified a broad component with FWHM$\sim$2000 km
s$^{-1}$ at the basis of its narrow \ha.  From Della Ceca et
al. (2001) we know that the luminosity of such broad component is
estimated by the above authors to be $L_{{\rm broad\;
H\alpha}}=8.0\times10^{40}$ erg s$^{-1}$. Assuming the 2-10~keV X-ray
luminosity by Della Ceca et al. (2001) scaled to our adopted distance
($L_X=6.1\times10^{42}$ erg s$^{-1}$) NGC~7679 falls among the
dust-unabsorbed Type~1 AGNs with a underluminous broad \ha\ component,
as shown in Fig.~\ref{fig:imanishi}. This is strictly agreement with
the proposed classification by Della Ceca et al. (2001).

%% Fig.9

At the same time NGC~7679 does offer {\em much more} to the
researchers currently exploring the elusive link between AGN activity
and ongoing SB. For instance, hosting simultaneously a bar and having
underwent a past interaction, it has at disposal at least two,
distinct efficient mechanisms to drive gas from its outer parts to the
nuclear region ({\em cf.} Gonz\`alez Delgado et al. 1998). Once the
expected inward flow of gas reaches its central region, is capable not
only of triggering star formation, but also of forming a gas reservoir
(in the form of a molecular torus) which likely feeds the nucleus
itself. This twofold scenario, is further enriched by the presence of
an amount of molecular hydrogen ---as derived form radio CO
observations (Kandalyan 2003)--- comparable to that of neutral
hydrogen (a few 10$^9$ M$_\odot$).  What is more, unlike other
Markarian galaxies, the CO in NGC~7679 distinguishes itself for having
the line FWHM larger than that of its \hi\ line.  Since the CO gas
kinematics reveals the rotation and/or velocity dispersion close to
the nucleus, this phenomenon is interpreted as due to a rapidly
rotating nuclear disk (cf. Kandalyan 2003). As such, part of this
circumnuclear gas is likely prone to stream inward with velocities
$\sim$ 100 km s$^{-1}$ directly onto the nucleus ({\em cf. } Regan et
al. 1999), so that a phenomenon usually unobservable in other AGN,
ends up showing itself largely ``in the open'' at the center of this
galaxy.

\section{Summary and conclusions}

NGC~7679 is a very attractive galaxy, where a wealth of distinct
{\em interlaced} phenomena do coexist.
The
investigation of its special properties is made easier both by its
relative proximity and projected orientation.

Among the most interesting features we were able to identify are the
following:

\begin{enumerate}

\item High resolution \ha\ imaging shows several knots and loose
      ionized gas debris. In addition it reveals a central
      circumnuclear star-forming spiral/ring capable of producing
      $\sim$75\% of the optical line emission within a radius of
      $\sim$1~kpc. Observations from literature suggest that this
      central regions contains contains also a large amount of
      molecular gas, likely giving rise to a stream moving inward, so
      as to possibly fuel the galaxy nucleus.

\item A ionized gas ring-like feature is part of the centrally located
      starburst dominating the galaxy emission both at optical {\em
      and} IR wavelengths. Such an event is characterized by a SFR of
      $\sim$10 M$_\odot$ yr$^{-1}$, as derived from optical and UV
      observations, and consistent also with 14 GHz radio continuum
      observations. The onset of the phenomenon is not earlier than
      20~Myr ago, as shown by the prominent stellar wind absorption
      lines seen in its IUE ultraviolet spectrum shortward of
      2000~\AA.

\item The ionized gas velocity field can be traced out to $r\sim20''$
      (corresponding to 5-6~kpc). Though the latter appears
      characterized by a basically regular rotation within the
      innermost ($15''$) region, clear signs of asymmetries ({\em
      i.e.} perturbations) are recognizable at a radius just beyond
      such a distance.

\item The overall neutral gas velocity field (as derived from \hi\
      radio observations), is perhaps more indicative of the past
      history of the binary system. In particular, although reflecting
      the combined outer kinematics of {\em both} NGC~7679 and
      NGC~7682, still allows observers to isolate their individual
      kinematic properties. In this respect, the most outstanding
      outcome is the macroscopic misalignment of the \hi\ and optical
      kinematic major axes of the companion galaxy NGC~7682
      ($\Delta$PA$>50^\circ$).

\item The extent of the neutral gas common envelope, in turn, suggests
      that a close encounter likely happened a few hundreds million
      years ago. Its relationship with the current, much more recent,
      central starburst has to be further investigated.

\item We unambiguously established that the angularly close fainter
      companion galaxy of NGC~7679, overposed by chance on its highly
      disturbed eastern arm, is indeed an object located much farther
      away, in the deep background of the main galaxy and, as such,
      certainly not involved in its dynamical evolution.

\item Della Ceca et al. (2001) are induced to look at NGC~7679 as a highly peculiar object
      owing to its {\em selective} absorption apparently not affecting
      its X-ray AGN emission, so as to ascribe to its nucleus a
      ``unique'' physics/geometry. Conversely, the present analysis
      leads to a {\em less extreme} classification of NGC~7679 {\em
      i.e.} as a ``normal'' low-luminosity Seyfert~1 object (given the
      ratio of its broad H$\alpha$ luminosity {\em vs.} hard X
      luminosity, {\em cf}. Fig.~\ref{fig:imanishi}) and a transition
      object at optical wavelengths ({\em cf}.
      Fig.~\ref{fig:kauffmann}). This allows us to conclude that
      NGC~7679 is basically a galaxy where starburst and a mildly
      obscured AGN (given the upper limit of the column density of the
      neutral hydrogen corresponding to E(B-V)$\sim$0.1 according to
      the above authors) cohexist with the starburst dominating
      the bolometric luminosity.
\end{enumerate}

\begin{acknowledgements}

MC acknowledges support from a VENI grant 639.041.203 awarded
by the Netherlands Organization for Scientific Research (NWO).

\end{acknowledgements}

\end{document}